\documentclass[jmp, amsmath, amssymb, ppreprint]{revtex4-1}

\usepackage[hyperindex, breaklinks, colorlinks=true, linkcolor={blue}, citecolor={blue},urlcolor={blue}]{hyperref} 

\usepackage{bm} 
\usepackage{mathtools} 

\newcommand{\R}{\ensuremath{\mathbf{R}}} 
\newcommand{\C}{\ensuremath{\mathbf{C}}} 
\newcommand{\Z}{\ensuremath{\mathbf{Z}}} 
\newcommand{\M}{\ensuremath{\mathcal{M}}} 
\newcommand{\LLambda}{\boldsymbol{\Lambda}} 
\newcommand{\U}{\ensuremath{\mathbf{U}}} 
\newcommand{\SO}{\text{SO}} 
\newcommand{\neref}[1]{(\ref{#1})} 
\newcommand{\eref}[1]{Eq.~(\ref{#1})} 
\newcommand{\erefs}[2]{Eqs.~(\ref{#1}) and (\ref{#2})} 
\newcommand{\nerefs}[2]{(\ref{#1}) and (\ref{#2})} 
\newcommand{\rref}[1]{Ref.~\cite{#1}} 
\newcommand{\rrefs}[2]{Refs.~\cite{#1} and \cite{#2}} 
\renewcommand{\S}{\mathfrak{S}} 
\newcommand{\Sphere}{\mathbf{S}} 
\newcommand{\defeq}{\coloneqq} 
\renewcommand{\d}{\text{d}} 
\renewcommand{\vec}[1]{\boldsymbol{#1}} 

\begin{document}

\title{Multipole Hair of Schwarzschild-Tangherlini Black Holes}

\author{Matthew S. Fox}

\email{msfox@g.hmc.edu}

\affiliation{Department of Physics, Harvey Mudd College, Claremont, CA 91711, USA}

\date{\today}

\begin{abstract}
We study the field of an electric point charge that is slowly lowered into an $n + 1$ dimensional Schwarzschild-Tangherlini black hole. We find that if $n > 3$, then countably infinite nonzero multipole moments manifest to observers outside the event horizon as the charge falls in. This suggests the final state of the black hole is not characterized by a Reissner-Nordstr\"om-Tangherlini geometry. Instead, for odd $n$, the final state either possesses a degenerate horizon, undergoes a discontinuous topological transformation during the infall of the charge, or both. For even $n$, the final state is not guaranteed to be asymptotically-flat.
\end{abstract}


\maketitle

\section{Introduction}
\label{sec:Intro}

The properties of four-dimensional black holes are rigidly constrained. For instance, all stationary and asymptotically-flat black hole solutions to the Einstein-Maxwell equations are topologically spherical and unique up to the choice of three asymptotic observables: mass, electric charge, and angular momentum \cite{Israel1, Israel2, Carter1, Carter2, Hawking, Robinson, Heusler}. This is Wheeler's famous ``no-hair theorem'' \cite{Wheeler}.

Higher dimensional black holes are less constrained, largely for two reasons (see \rref{Emparan0} for a separate and less heuristic perspective). One, the rotation group $\SO(n)$ permits $\left\lfloor \frac{n}{2} \right\rfloor$ independent angular momenta. Accordingly, the rotational degrees of freedom of black holes in $n + 1$ dimensional spacetime become progressively more complex as $n$ increases \cite{Hollands, Emparan1}. Furthermore, black holes with fixed masses in $n \geq 5$ spatial dimensions may have arbitrarily large angular momentum \cite{Myers}. Two, Hawking's theorem on the topology of black holes \cite{Hawking} does not directly generalize to higher dimensions because his proof relies on the Gauss-Bonnett theorem. Although topological restrictions exist for higher dimensional black holes \cite{Galloway1, Galloway2, Helfgott}, a hyperspherical topology is not the only option \cite{Emparan1, Costa}. As a result, extended black $p$-branes are not precluded in higher dimensional spacetimes \cite{Emparan1, Emparan2, Costa}. These results imply that the uniqueness theorems for four-dimensional black holes do not immediately extend to higher dimensions.

However, if restricted to solutions with hyperspherical topology and non-degenerate horizons, then the Schwarzschild-Tangherlini (ST) black hole \cite{Tangherlini} is the unique static and asymptotically-flat vacuum solution to the higher dimensional Einstein equations \cite{Hollands, Hwang, Gibbons1, Gibbons2}. Furthermore, the higher dimensional Reissner-Nordstr\"om (RN-Tangherlini, or simply RNT) black hole is the unique static and asymptotically-flat electrovac solution to the higher dimensional Einstein-Maxwell equations \cite{Gibbons3, Ida}. Non-uniqueness is most apparent in the context of stationary black hole solutions \cite{Emparan1, Costa, Myers, Emparan2}. 

For four-dimensional black holes, the no-hair theorem implies that the process of slowly \footnote{By ``slowly'' we mean ``slow enough that our static considerations remain valid''.} lowering an electric point charge of strength $q$ into a Schwarzschild black hole of mass $M$ results in a RN black hole of mass $M$ and charge $q$. Furthermore, the resulting black hole does not possess unconserved charges like electric multipole moments (excluding the monopole) as these are ``hair'' for the black hole. The details of this process can be found in \rref{Cohen}.

That in four dimensions the slow infall of an electric charge into a Schwarzschild black hole results in only one type of black hole --- the RN black hole --- may be viewed as a corollary of the uniqueness theorem for RN black holes. In the same way, the uniqueness of RNT black holes ostensibly implies that a sufficiently slow infall of an electric charge into a ST black hole results in a unique final state --- the RNT black hole. If this is the final state, then, due to the hyperspherical symmetry of RNT spacetime, all electric multipole moments (except the monopole) necessarily vanish as the charge approaches the event horizon.

Following the analyses of \rrefs{Cohen}{Persides}, we prove the contrary: if an electric point charge falls slowly into a ST black hole, then the final state acquires countably infinite nonzero multipole moments. Depending on the spatial dimension $n$, these multipole fields need not even be finite. This suggests the resulting black hole is not RNT in nature, and, depending on $n$, brings about the possibility of destruction of the horizon.

In this paper, we employ the metric signature $(- + \cdots +)$ and work in the natural system of units in which $c = G = 1$. We also adopt the following notation: $\S^{n+1}$ is $n+1$ dimensional ST spacetime, $\R^n$ is $n$ dimensional Euclidean space, $\C$ is the complex plane, $\Z^+$ is the set of positive integers, $\Z^* \defeq \Z^+ \cup \{0\}$ is the set of nonnegative integers, and $\Sphere^{n-1}$ is the unit $n - 1$ sphere.

\section{Schwarzschild-Tangherlini Geometry}
\label{sec:STGeometry}

The $n+1$ dimensional ST black hole is described by the $n+1$ dimensional ST spacetime, $\S^{n+1}$. In this spacetime, there exists a chart $(\U^{n+1}, \vec{\psi})$  (the \emph{ST chart}) with map $\vec{\psi} \defeq (t,r, \vec{\varphi}) : \U^{n+1} \subseteq \S^{n+1} \rightarrow \R^{n+1}$ that reduces to the canonical four-dimensional Schwarzschild map $(t,r,\theta, \phi) : \U^{3+1} \subseteq \S^{3+1} \rightarrow \R^4$ when $n = 4$. In this way, the ST chart is a dimensional continuation of the four-dimensional Schwarzschild chart. In the ST chart, the coordinates $t : \U^{n+1} \rightarrow \R$ and $r : \U^{n+1} \rightarrow \R$ retain the meaning (outside the event horizon) of ``time as measured by an asymptotic observer'' and ``circumferential radius as measured by an asymptotic observer,'' respectively. The angular functions $(\theta, \phi)$, however, are generalized to the hyperspherical coordinates $\vec{\varphi} \defeq (\varphi_1,\dots \varphi_{n-1})$, where $\varphi_i : \U^{n+1} \rightarrow [0, \pi]$ for $i = 1,\dots, n-2$ and $\varphi_{n-1}: \U^{n+1} \rightarrow [0,2\pi)$.

In the ST chart, the metric $g$ of $\S^{n+1}$ possesses the line element \cite{Tangherlini}
\begin{equation}
g(\d\vec{\psi},\d\vec{\psi}) = -\left(1 - \frac{2\M}{r^{n-2}}\right)\d t^2 + \left(1 - \frac{2\M}{r^{n-2}}\right)^{-1}\d r^2 + r^2\gamma(\d\vec{\varphi}, \d\vec{\varphi}),
\label{eq:SchwarzschildMetricHigherDimensions}
\end{equation}
where $\gamma$ is the metric of $\Sphere^{n-1}$ with line element
\begin{equation}
\gamma(\d\vec{\varphi}, \d\vec{\varphi}) = \d\varphi^2_1 + \sum_{i=2}^{n-1}\prod_{j=1}^{i-1} \sin^2\varphi_j\, \d\varphi_i^2.
\label{eq:n-1sphere}
\end{equation}
The value $\M$ in \eref{eq:SchwarzschildMetricHigherDimensions} is a constant related to the physical mass $M$ of the black hole by
\begin{equation}
\M \defeq \frac{8\pi M}{(n-1)\Omega_{n-1}},
\label{eq:massparameter}
\end{equation}
where $\Omega_{n-1} \defeq 2\pi^{n/2} / \Gamma(n/2)$ is the volume of $\Sphere^{n-1}$. The singular nature of \eref{eq:SchwarzschildMetricHigherDimensions} at $r = r_s \defeq (2\M)^{1/(n-2)}$ (the \emph{ST radius}) is an artifact of the choice of chart (an appropriate diffeomorphism will transform it away). However, the singular nature at $r = 0$ is a genuine curvature singularity (the Kretschmann scalar is infinite there). The locus of points for which $r = r_s$ constitute the event horizon of the black hole and the singular point for which $r = 0$ is the singularity.

Of interest to us is the effect of the geometry \neref{eq:SchwarzschildMetricHigherDimensions} on the form of Laplace's equation. Using the abstract index notation, the Laplacian $\Delta$ on a general $n+1$ dimensional spacetime with metric $g$ is defined by \cite{Thorne}
\begin{equation}
\Delta \defeq \frac{1}{\sqrt{|\det g|}}\partial_i\left(\sqrt{|\det g|}g^{ij}\partial_j\right),
\label{eq:LaplaceOperator}
\end{equation}
where vertical bars $|\cdot|$ denote absolute value and Latin indices run over the spatial components. In the ST chart, Latin indices will run from $1$ to $n$ corresponding to $r,\varphi_1, \dots, \varphi_{n-1}$, respectively. The Laplacian on $\S^{n+1}$ in the ST chart, $\Delta_{\S^{n+1}}$, is thus
\begin{equation}
\Delta_{\S^{n+1}} = \frac{1}{r^{n-1}} \partial_r\left[r^{n-1}\left(1 - \frac{r_s^{n-2}}{r^{n-2}}\right)\partial_r\right] + \frac{1}{r^2}\Delta_{\Sphere^{n-1}},
\label{eq:LaplaceSchwarzschild}
\end{equation}
where $\Delta_{\Sphere^{n-1}}$ is the Laplacian on $\Sphere^{n-1}$ in hyperspherical coordinates (the \emph{hyperspherical Laplacian}). 

\subsection{Hyperspherical Harmonics}
\label{sec:Hyperharmonics}

The eigenfunctions of the hyperspherical Laplacian constitute the higher dimensional generalization of the canonical spherical harmonics on $\Sphere^2$. These eigenfunctions are the \emph{hyperspherical harmonics.} Specifically, an \emph{$n$ dimensional hyperspherical harmonic of degree $k \in \Z^*$} is a map $Y_k:\Sphere^{n-1} \rightarrow \C$ satisfying
\begin{equation}
\Delta_{\Sphere^{n-1}} Y_k(\vec{\varphi}) = -k(k + n - 2)Y_k(\vec{\varphi}),
\label{eq:Eigenfunction}
\end{equation}
among other conditions \cite{Frye}. Indeed, for the case $n = 3$, \eref{eq:Eigenfunction} reduces to the equation $\Delta_{\Sphere^2} Y_k(\vec{\varphi}) = -k(k+1) Y_k(\vec{\varphi})$, which is familiar from quantum mechanics.

Importantly, if $k \neq l$, then the functions $Y_k(\vec{\varphi})$ and $Y_l(\vec{\varphi})$ can be chosen to be orthogonal over $\Sphere^{n-1}$ with respect to the inner product \cite{Wen}
\begin{equation}
\langle Y_k, Y_l\rangle \defeq \int_{\Sphere^{n-1}} \hat{Y}_k(\vec{\varphi}) Y_l(\vec{\varphi})\, \d \vec{\Omega}_{n-1} = 0,
\label{eq:innerProduct}
\end{equation}
where a hat denotes complex conjugation and $\d \vec{\Omega}_{n-1} \defeq \sqrt{\det \gamma}\, \d \vec{\varphi}$ is the natural volume form on $\Sphere^{n-1}$.

For fixed $n \geq 3$, the degree of a hyperspherical harmonic completely determines the number of hyperspherical harmonics of the same degree that are linearly independent to it. With this in mind, we denote by $\Gamma_{k}$ the number of linearly independent hyperspherical harmonics of degree $k$. For $n \geq 3$, a combinatorial argument \cite{Frye, Wen} proves
\begin{equation}
\Gamma_k = \frac{(2k + n-2)(n + k - 3)!}{k!(n-2)!}.
\label{eq:GammaExplicitForm}
\end{equation}
The Gram-Schmidt orthonormalization procedure then allows one to produce an orthonormal set of hyperspherical harmonics $\{Y_k^i\}_{i = 1}^{\Gamma_{k}}$ that satisfy
\begin{equation}
\langle Y_k^i, Y_l^j\rangle = \int_{\Sphere^{n-1}} \hat{Y}_k^{i}(\vec{\varphi}) Y_l^j(\vec{\varphi})\, \d \vec{\Omega}_{n-1} = \delta_{k,l}\delta_{i,j},
\label{eq:innerProductTwo}
\end{equation}
where $\delta_{k,l}$ is the Kronecker delta. These functions constitute an orthonormal basis for all square-integrable functions on $\Sphere^{n-1}$ \cite{Frye}. Thus, the hyperspherical harmonics obey the completeness relation
\begin{equation}
\sum_{k \geq 0}\sum_{l = 1}^{\Gamma_{k}} \hat{Y}_k^{l}(\vec{\vartheta})Y_k^l(\vec{\varphi}) = \delta^{n-1}(\vec{\varphi} - \vec{\vartheta}),
\label{eq:completeness}
\end{equation}
where $\vec{\vartheta}: \U^{n+1} \rightarrow \Sphere^{n-1}$ is a hyperspherical coordinate and $\delta$ is the Dirac delta function.

\subsection{Poisson's Equation}
\label{sec:Poisson}

Consider now a real-valued test field $\Psi: \U^{n+1} \rightarrow \R$, i.e., a real scalar field weak enough that the geometry is unaffected by it. Let $\Psi$ satisfy the d'Alembert wave equation,
\begin{equation}
\Box \Psi \defeq \nabla^\mu\nabla_\mu \Psi = \Omega_{n-1} f(t, r, \vec{\varphi}),
\label{eq:WaveEquation}
\end{equation}
where $f:\U^{n+1} \rightarrow \R$ is a well-behaved function, $\nabla$ is the covariant derivative with respect to the metric $g$ on $\S^{n+1}$, and Greek indices run from $0$ to $n$ corresponding to $t,r, \varphi_1,\dots, \varphi_{n-1}$, respectively. In the case where both $\Psi$ and $f$ are time-independent, \eref{eq:WaveEquation} reduces to Poisson's equation,
\begin{equation}
\Delta_{\S^{n+1}} \Psi(r,\vec{\varphi}) = \Omega_{n-1} f(r,\vec{\varphi}).
\label{eq:Poisson}
\end{equation}
We now show that the equation of motion for the field $\Psi$ of an electrostatic charge in the vicinity of a ST black hole satisfies Poisson's equation. 

In a general curved spacetime with metric $g$, Maxwell's equations can be written as \cite{Thorne}
\begin{equation}
\Omega_{n-1}j^\nu = \frac{1}{\sqrt{|\det g|}}\partial_\mu\left(\sqrt{|\det g|}F^{\mu\nu}\right),
\label{eq:Maxwell}
\end{equation}
where $F_{\mu\nu} \defeq \partial_\mu A_\nu - \partial_\nu A_\mu$ are the components of the Faraday tensor and $A_\mu$ are the components of the electromagnetic potential. In particular, in an appropriate gauge, $\Psi \defeq A_0$ is the electric potential. Assuming no magnetic fields ($A_i = 0$) and static electric fields (time-independent $\Psi$), the vector current $j^\nu$ vanishes trivially for $\nu = 1,\dots, n+1$. However, for $\nu = 0$, \eref{eq:Maxwell} reduces to the nontrivial expression
\begin{equation}
\Omega_{n-1} j^0 = g^{00} \Delta_{\S^{n+1}}\Psi - \frac{g^{00}r_s^{n-2}(n-2)}{r^{n-1}}\partial_r \Psi.
\label{eq:MaxwellReduction}
\end{equation}
As $\Psi(r,\vec{\varphi})$ is time-independent (by assumption), so is the \emph{physical source} $j^0(r,\vec{\varphi})$. Hence, the electrostatic problem reduces to solving Poisson's equation with the \emph{effective source}
\begin{equation}
f_{\text{eff}}(r, \vec{\varphi}) \defeq g_{00}j^0(r,\vec{\varphi}) + \frac{r_s^{n-2}(n-2)}{\Omega_{n-1}r^{n-1}}\partial_r \Psi(r,\vec{\varphi}).
\label{eq:effectiveSource}
\end{equation}
Consequently, to understand \eref{eq:Poisson} is to understand electrostatics in $\S^{n+1}$.

\section{Radial Equation of Motion}
\label{sec:Radialequation}

Consider Poisson's equation \neref{eq:Poisson} with the effective source \neref{eq:effectiveSource}, but in the absence of physical sources, $j^0(r, \vec{\varphi}) = 0$. Then, \eref{eq:Poisson} reduces to
\begin{equation}
\Delta_{\S^{n+1}} \Psi(r,\vec{\varphi}) = \frac{r_s^{n-2}(n-2)}{r^{n-1}}\partial_r \Psi(r,\vec{\varphi}).
\label{eq:Laplace}
\end{equation}
We look for solutions to \eref{eq:Laplace} of the form 
\begin{equation}
\Psi(r, \vec{\varphi}) = \sum_{k \geq 0} R_k(r)Y_k(\vec{\varphi}).
\label{eq:separationofvariables}
\end{equation}
Using \eref{eq:LaplaceSchwarzschild} and the eigenfunction relation \neref{eq:Eigenfunction}, one deduces that, for each $k \in \Z^*$, $R_k(r)$ must satisfy
\begin{equation}
R_k'' + \frac{(n-1)(r^{n-2} - r_s^{n-2})}{r(r^{n-2} - r_s^{n-2})}R_k' - \frac{k(k+n-2)r^{n-4}}{r^{n-2} - r_s^{n-2}}R_k = 0.
\label{eq:SeparatedVariablesGeneral}
\end{equation}
For later convenience, we abbreviate the polynomial coefficients to
\begin{equation}
P_k(r, r_s) \defeq \frac{(n-1)(r^{n-2} - r_s^{n-2})}{r(r^{n-2} - r_s^{n-2})}
\label{eq:coefficientOne}
\end{equation}
and
\begin{equation}
Q_k(r,r_s) \defeq -\frac{k(k+n-2)r^{n-4}}{r^{n-2} - r_s^{n-2}}.
\label{eq:coefficientTwo}
\end{equation}
Note that the differential equation \neref{eq:SeparatedVariablesGeneral} is invariant under the exchange $k \leftrightarrow -(k + n - 2)$. Hence, given a solution, a second solution follows by swapping $k \leftrightarrow -(k + n-2)$. Of course, one must check that this second solution is linearly independent of the first.

The differential equation \neref{eq:SeparatedVariablesGeneral} has three nonessential singularities at $r = 0, r = r_s,$ and $r = \infty$. When $r_s = 0$, the singularities are $r = 0$ and $r = \infty$, and two independent solutions are $r^k$ and $r^{-(k+n-2)}$. Evidently, these solutions are valid for all $r \in (0, \infty)$. When $r_s \neq 0$, we substitute $\rho \defeq (\frac{r_s}{r})^{n-2}$ into \eref{eq:SeparatedVariablesGeneral}, which becomes
\begin{equation}
\rho^2(\rho - 1)\ddot{R}_k + \frac{k(k+n-2)}{(n-2)^2}R_k = 0,
\label{eq:radialEquation}
\end{equation}
where a dot denotes differentiation with respect to $\rho$. This is a special case of the hypergeometric differential equation. The equation has three nonessential singularities at $\rho = 0$, $\rho = 1$, and $\rho = \infty$ corresponding to $r = \infty$, $r = r_s,$ and $r = 0$, respectively. 

We shall solve the differential equation \neref{eq:radialEquation} around $\rho = 0$ for two reasons. One, after transitioning back to the ST chart, Frobenius' method \cite{Nagle} guarantees a convergent solution for all $r \in (r_s, +\infty)$, which is the desired region of study. Two, only by solving around $\rho = 0$ is the physically meaningful limit $r_s \rightarrow 0^+$ [$M\rightarrow 0^+$ in \eref{eq:massparameter}] well-defined in the solution. To understand the second point, first note that the structure of \eref{eq:SeparatedVariablesGeneral} is such that
\begin{equation}
\lim_{r_s\rightarrow 0^+}\lim_{r \rightarrow \infty} rP_k(r,r_s) = \lim_{r \rightarrow \infty}\lim_{r_s\rightarrow 0^+} rP_k(r,r_s)
\label{eq:commutingLimits}
\end{equation}
and
\begin{equation}
\lim_{r_s\rightarrow 0^+}\lim_{r \rightarrow \infty} r^2Q_k(r,r_s) = \lim_{r \rightarrow \infty}\lim_{r_s\rightarrow 0^+} r^2Q_k(r,r_s)
\label{eq:commutingLimitsTwo}
\end{equation}
for all $k\in \Z^*$. That these two pairs of limits commute implies the indicial equation around $r = +\infty$ for \eref{eq:SeparatedVariablesGeneral} does not change as $r_s \rightarrow 0^+$. Accordingly, the form of the solutions is the same for all $r_s \geq 0$. This is obviously crucial if the solutions to \eref{eq:SeparatedVariablesGeneral} with $r_s \neq 0$ are to reduce to $r^k$ and $r^{-(k + n-2)}$ in the limit as $r_s\rightarrow 0^+$. Incidentally, this does not happen if the differential equation is solved around either $r = 0$ or $r = r_s$.

The differential equation \neref{eq:radialEquation} is solved by first noting that around $\rho = 0$, the indicial equation has roots
\begin{equation}
\chi^{\pm}_k \defeq \frac{1\pm 1}{2} + \frac{k}{n-2}.
\label{eq:indicialRoots}
\end{equation}
Clearly, $\chi^+_k > \chi^-_k$ and $\chi^+_k - \chi^-_k \in \Z^*$ if and only if $(n-2) \mid k$, where $\mid$ means ``divides.'' One solution to \eref{eq:radialEquation} is then of the form
\begin{equation}
R_k^{(\alpha)}(r,r_s) \defeq \sum_{m \geq 0} \alpha_{k, m} \rho^{m + \chi^+_k},
\label{eq:SeriesFormOne}
\end{equation}
and a second follows from the exchange $k \leftrightarrow -(k+n-2)$,
\begin{equation}
R^{(\bar{\alpha})}_k(r,r_s) \defeq \sum_{m \geq 0} \bar{\alpha}_{k, m} \rho^{m - \chi^-_k}.
\label{eq:SeriesFormTwo}
\end{equation}
Here, we have utilized the relation
\begin{equation}
\chi_{-(k + n-2)}^\pm = -\chi^\mp_k.
\label{eq:chiRelation}
\end{equation}
In \erefs{eq:SeriesFormOne}{eq:SeriesFormTwo}, $\{\alpha_{k,m}\}_{m\geq0}$ and $\{\bar{\alpha}_{k,m}\}_{m \geq 0}$ are $k$- and $m$-dependent sequences of real numbers for which $\alpha_{k,0} \neq 0$ and $\bar{\alpha}_{k,0} \neq 0$ for all $k \in \Z^*$. The barred sequence is related to the unbarred sequence via the conjugation $k \leftrightarrow -(k+n-2)$, i.e., $\bar{\alpha}_{k,m}\defeq \alpha_{-(k + n-2),m}$. Substituting \eref{eq:SeriesFormOne} into \eref{eq:radialEquation} establishes that each $\alpha_{k,m}$ must satisfy
\begin{equation}
\tilde{\alpha}_{k,m} \defeq \frac{\alpha_{k, m}}{\alpha_{k,0}} = \frac{\left(\chi^+_k\right)_m \left(\chi^-_k\right)_m}{m!\left(2\chi^+_k\right)_m},
\label{eq:sequenceAlpha}
\end{equation}
where $(x)_m \defeq x(x + 1)\cdots(x + m - 1)$ is the Pochhammer symbol defined such that $(x)_0 = 1$ for all $x \in \R$. Accordingly, from \eref{eq:chiRelation}, the barred sequence satisfies
\begin{equation}
\tilde{\bar{\alpha}}_{k,m} \defeq \frac{\bar{\alpha}_{k, m}}{\bar{\alpha}_{k,0}} = \frac{\left(-\chi^-_k\right)_m \left(- \chi^+_k\right)_m}{m!\left(-2\chi^-_k\right)_m}.
\label{eq:sequenceAlphaBar}
\end{equation}
Importantly, the unbarred sequence $\{\alpha_{k,m}\}_{m\geq0}$ terminates if and only if $k = 0$, while the barred sequence $\{\bar{\alpha}_{k,m}\}_{m\geq0}$ terminates if and only if there exists an $N_k \in \Z^*$ such that 
\begin{equation}
N_k = \min\left\{z \in \Z^*:\left(-\chi_k^-\right)_{z + 1} = 0 \quad \text{or} \quad \left(-\chi_k^+\right)_{z+1} = 0\right\}.
\label{eq:formulaForNk}
\end{equation}
Such an $N_k$ exists if and only if $(n-2) \mid k$, in which case \eref{eq:formulaForNk} implies $N_k = \chi_k^-$. With this in mind, we introduce the piecewise function
\begin{equation}
\Lambda_{k}\defeq 
\begin{cases}
\chi_k^- & \text{if $(n-2) \mid k$},\\
+\infty & \text{otherwise}.
\end{cases}
\label{eq:LambdaFunction}
\end{equation}
Two solutions to \eref{eq:SeparatedVariablesGeneral} are then
\begin{equation}
R_k^{(\alpha)}(r,r_s) = r_s^{-(k + n - 2)} \sum_{m \geq 0} \tilde{\alpha}_{k,m}\left(\frac{r_s}{r}\right)^{k + (m+1)(n-2)}
\label{eq:seriesSolutionOneComplete}
\end{equation}
and
\begin{equation}
R_k^{(\bar{\alpha})}(r,r_s) = r_s^{k} \sum_{m = 0}^{\Lambda_{k}} \tilde{\bar{\alpha}}_{k,m}\left(\frac{r}{r_s}\right)^{k - m(n-2)}.
\label{eq:seriesSolutionTwoComplete}
\end{equation}
Here, we have fixed $\alpha_{k,0} = r_s^{-(k + n - 2)}$ and $\bar{\alpha}_{k,0} = r_s^k$ so that $R^{(\alpha)}_k \rightarrow r^{-(k + n - 2)}$ and $R^{(\bar{\alpha})}_k \rightarrow r^{k}$ as $r_s \rightarrow 0^+$, as desired. Crucially, both solutions are absolutely convergent on $(r_s,+\infty)$ by Frobenius' method \cite{Nagle}. The linear independence of the solutions follows from the fact that, asymptotically,
\begin{equation}
R_k^{(\alpha)}(r,r_s) \sim r^{-(k + n-2)}
\label{eq:asymptoticOne}
\end{equation}
and
\begin{equation}
R_k^{(\bar{\alpha})}(r,r_s) \sim r^{k}.
\label{eq:asymptoticTwo}
\end{equation}
Thus, assuming $n \geq 3$, the Wronskian,
\begin{equation}
W\left[R^{(\alpha)}_k, R^{(\bar{\alpha})}_k\right] \defeq R^{(\alpha)}_k R'^{(\bar{\alpha})}_k - R'^{(\alpha)}_k R^{(\bar{\alpha})}_k \sim \frac{2k + n-2}{r^{n-1}},
\label{eq:WronskianDefinition}
\end{equation}
is nonzero asymptotically for all $k\in \Z^*$. Hence, for $n \geq 3$, the Wronskian is nonzero on $(r_s, +\infty)$ for all $k\in \Z^*$, so the solutions \nerefs{eq:seriesSolutionOneComplete}{eq:seriesSolutionTwoComplete} are linearly independent on $(r_s, +\infty)$ for all $k\in \Z^*$. Abel's identity \cite{Nagle} proves that the general, non-asymptotic Wronskian is the same as the asymptotic value given in \eref{eq:WronskianDefinition}. Therefore,
\begin{equation}
W\left[R^{(\alpha)}_k, R^{(\bar{\alpha})}_k\right] = \frac{2k + n-2}{r^{n-1}}.
\label{eq:Wronskian}
\end{equation}
Evidently, the Wronskian is nonzero and finite for all $r \in (r_s, +\infty)$, and is likewise ($ = \frac{2k + n-2}{r^{n-1}_s}$) as $r\rightarrow r_s^+$. These statements hold true for all $k \in \Z^*$ when $n\geq 3$. We now study the behavior of the solutions \nerefs{eq:seriesSolutionOneComplete}{eq:seriesSolutionTwoComplete} as $r\rightarrow r_s^+$.

Since these solutions are Gaussian hypergeometric functions, Gauss' hypergeometric theorem \cite{Bailey} proves
\begin{equation}
\lim_{r \rightarrow r_s^+}R_k^{(\alpha)}(r,r_s) = \frac{\Gamma(2\chi_k^+)}{\Gamma(\chi^+_k)\Gamma\left(1 + \chi_k^+\right)}r_s^{-(k+n-2)}.
\label{eq:schwarzschildTwo}
\end{equation}
This limit converges for all $k \geq 0$ when $n \geq 3$. On the other hand,
\begin{equation}
\lim_{r \rightarrow r_s^+}R_k^{(\bar{\alpha})}(r,r_s) =
r_s^k\begin{cases}
\sum_{m = 0}^{\Lambda_{k}}\tilde{\bar{\alpha}}_{k,m}& \text{if $(n-2) \mid k$},\\
\frac{\Gamma(-2\chi^-_k)}{\Gamma(-\chi^-_k)\Gamma\left(1 - \chi_k^-\right)} & \text{otherwise}.
\end{cases}
\label{eq:schwarzschildOne}
\end{equation}
We study this limit in the two possible cases.

First, suppose $(n-2) \nmid k$. Then, $n > 3$ \footnote{Otherwise ($n = 3$), $(n-2) \mid k$ for all $k \in \Z^*$, so the case we are considering never applies.}, $\Lambda_k = +\infty,$ and $\chi^-_k \defeq \frac{k}{n-2} \not\in \Z^*$. The analytically continued gamma function has a simple pole at each nonpositive integer. Thus, \eref{eq:schwarzschildOne} converges if and only if $2\chi_k^- \not\in \Z^+$. Since $(n-2) \nmid k$, $2\chi_k^{-} \not\in \Z^+$ if and only if $n \neq 2(d_k + 1)$, where $d_k \in \Z^+$ is a divisor of $k$.

Now suppose $(n-2) \mid k$. Then, $\Lambda_k = \chi_k^- \in \Z^*$. As a result, \eref{eq:schwarzschildOne} converges for all $k \in \Z^*$ when $n \geq 3$. In fact, if $k \neq 0$, then
\begin{equation}
\sum_{m = 0}^{\Lambda_k} \tilde{\bar{\alpha}}_{k,m} = -\frac{(-\Lambda_k - 1)_{\Lambda_k+1} (-\Lambda_k)_{\Lambda_k+1}}{(\Lambda_k+1)!(-2\Lambda_k)_{\Lambda_k+1}} = 0,
\label{eq:vanishingSum}
\end{equation}
where equality to zero follows from $(-\Lambda_k)_{\Lambda_k+1} = 0$. Using $R_0^{(\bar{\alpha})}(r_s, r_s) = 1$, \erefs{eq:schwarzschildOne}{eq:vanishingSum} imply
\begin{equation}
\lim_{r \rightarrow r_s^+}R_k^{(\bar{\alpha})}(r,r_s) =
\begin{cases}
1 & \text{if $k = 0$},\\
0 & \text{if $k \in \Z^+$ and $(n-2) \mid k$},\\
\frac{\Gamma(-2\chi^-_k)}{\Gamma(-\chi^-_k)\Gamma\left(1-\chi_k^-\right)}r_s^k & \text{otherwise}.\\
\end{cases}
\label{eq:vanishingOfEMMultipoles}
\end{equation}
These convergence and divergence properties constitute the origin of the electric multipole hair on ST black holes. They also indicate why $n = 3$ is special: only with this dimension does $(n-2) \mid k$, and thus is $R_k^{(\bar{\alpha})}(r,r_s) = 0$ as $r\rightarrow r_s^+$, for all $k \in \Z^+$.

Finally, we compute the derivative of $R_k^{(\alpha)}(r,r_s)$ as $r\rightarrow r_s^+$. The derivative properties of hypergeometric functions imply
\begin{equation}
\lim_{r\rightarrow r_s^+}R'^{(\alpha)}_k(r,r_s) = -(n-2)r_s^{-(k + n - 3)}\left[\tilde{\alpha}_{k,1}\sum_{m\geq0}\tilde{\beta}_{k,m} + \frac{\Gamma(2\chi_k^+)\chi^+_k}{\Gamma(\chi^+_k)\Gamma\left(1 + \chi_k^+\right)}\right],
\label{eq:derivativeOfField}
\end{equation}
where the sequence $\{\tilde{\beta}_{k,m}\}_{m\geq0}$ is defined by
\begin{equation}
\tilde{\beta}_{k,m} \defeq \frac{(1 + \chi_k^+)_m\left(\chi_k^+\right)_m}{m!(1 + 2\chi_k^+)_m}.
\label{eq:BetaSequence}
\end{equation}
The same methods used to obtain \erefs{eq:schwarzschildTwo}{eq:schwarzschildOne} show that the $\tilde{\beta}_{k,m}$ sum in \eref{eq:derivativeOfField} diverges for all $k \in \Z^*$. Consequently, $R'^{(\alpha)}_k(r,r_s)$ diverges as $r \rightarrow r_s^+$ for all $k \in \Z^*$ when $n\geq3$.

\section{Green's Function}
\label{sec:Greenfunction}

For sake of clarity, we write $\vec{r} \defeq (r,\vec{\varphi})$ and henceforth suppress all dependencies on $r_s$. We assume the physical source function $j^0(\vec{r})$ in \eref{eq:effectiveSource} is zero for sufficiently large $r$. Furthermore, we impose the Dirichlet boundary condition $G \sim r^{-(n-2)}$, where $G$ is the Green's function to be determined in this section. The situation we examine is when $j^0(\vec{r})$ is nonzero only at a singular point $p \in \S^{n+1}$, for which $\vec{r}(p) = \vec{r}_p \defeq (r_p, \vec{\varphi}_p)$ is constant. We shall assume $r_p > r_s$ until stated otherwise. In this case, the electric field is generated by a stationary point source outside the black hole. The source function is then a particular instance of the effective source \neref{eq:effectiveSource},
\begin{equation}
f_{\text{eff}}(\vec{r}) = g_{00}\delta^n(\vec{r} - \vec{r}_p) + \frac{r_s^{n-2}(n-2)}{\Omega_{n-1}r^{n-1}}\partial_r G(\vec{r}, \vec{r}_p).
\label{eq:PointSourceTwo}
\end{equation}
Here, the normalizations of the $\delta$ functions are chosen so that
\begin{equation}
\int \delta(r - r_p) r^{n-1}\, \d r = 1
\label{eq:normalizationOne}
\end{equation}
and
\begin{equation}
\int_{\Sphere^{n-1}}\delta^{n-1}(\vec{\varphi} - \vec{\varphi}_p)\, \d \vec{\Omega}_{n-1} = 1.
\label{eq:normalizationTwo}
\end{equation}
The solution to the Dirichlet problem is the Green's function $G(\vec{r}, \vec{r}_p)$. To find this, we first write
\begin{equation}
\delta^n(\vec{r} - \vec{r}_p) = \delta(r - r_p)\sum_{k \geq 0}\sum_{l = 1}^{\Gamma_{k}} \hat{Y}_k^{l}(\vec{\varphi}_p)Y_k^l(\vec{\varphi}),
\label{eq:alternativeDeltaFunction}
\end{equation}
where we have employed the completeness relation \neref{eq:completeness}. Next, we propose the ansatz
\begin{equation}
G(\vec{r}, \vec{r}_p) = \sum_{k\geq0}\sum_{l = 1}^{\Gamma_{k}}Z_k^l(\vec{\varphi}_p)R_k(r,r_p)Y_k^l(\vec{\varphi}).
\label{eq:Green'sFunctionForm}
\end{equation}
Here, $\{Z_k^l\}_{l=1}^{\Gamma_{k}}$ is a set of undecided, complex-valued functions on $\Sphere^{n-1}$. Substituting \eref{eq:Green'sFunctionForm} into \eref{eq:Poisson} establishes that $Z_k^l(\vec{\varphi}_p) = \hat{Y}_k^l(\vec{\varphi}_p)$ and that $R_k(r,r_p)$ satisfies 
\begin{equation}
\frac{d}{dr}\left[r^{n-1}\left(1 - \frac{r_s^{n-2}}{r^{n-2}}\right) R_k'\right] - r^{n-3}k(k+n-2)R_k - (n-2)r_s^{n-2}R_k' = g_{00} \Omega_{n-1}r^{n-1}\delta(r - r_p).
\label{eq:RadialEquationSatisfies}
\end{equation}
This is identical to \eref{eq:SeparatedVariablesGeneral} for all $r \neq r_p$. Therefore, $R_k(r,r_p)$ is a linear combination of both $R_k^{(\alpha)}$ and $R_k^{(\bar{\alpha})}$,
\begin{equation}
R_k(r,r_p) = 
\begin{cases}
A_kR_k^{(\alpha)}(r) + \bar{A}_kR_k^{(\bar{\alpha})}(r) & \text{if $r > r_p$},\\
B_kR_k^{(\alpha)}(r) + \bar{B}_kR_k^{(\bar{\alpha})}(r) & \text{if $r < r_p$}.
\end{cases}
\label{eq:generalForm}
\end{equation}
As $G \sim r^{-(n-2)},$ we require $\bar{A}_k = 0$ since $R^{(\bar{\alpha})}_k(r) \sim r^k$ by \eref{eq:asymptoticTwo}. We determine $B_k$ by requiring the Lorentz scalar
\begin{equation}
\frac{1}{2}F_{\mu\nu}F^{\mu\nu} = \left(\partial_r V\right)^2 + \left(1 - \frac{r_s^{n-2}}{r^{n-2}}\right)^{-1}(\partial_{\varphi_i}V)(\partial^{\varphi_i}V)
\label{eq:invariant}
\end{equation}
to be finite as $r \rightarrow r_s^+$ when $r_p > r_s$. As in \eref{eq:derivativeOfField}, $R'^{(\alpha)}_k$ is divergent for all $k\geq0$ as $r\rightarrow r_s^+$. We therefore set $B_k = 0$ to suppress the divergence of $\partial_rV$. Finally, we require that the solution be continuous at $r = r_p$. We conclude that
\begin{equation}
R_k(r,r_p) = C_kR_k^{(\alpha)}(r_>)R_k^{(\bar{\alpha})}(r_<),
\label{eq:radialFunctionForm}
\end{equation}
where $C_k \defeq A_k / R_k^{(\bar{\alpha})}(r_p) = \bar{B}_k / R_k^{(\alpha)}(r_p) $ is a constant and $r_< \defeq \min\{r,r_p\}$ while $r_> \defeq \max\{r,r_p\}$. At $r = r_p$, the function $R_k(r,r_p)$ is continuous (by design), though its first derivative is not. Integrating \eref{eq:RadialEquationSatisfies} over the interval $(r_p - \epsilon, r_p + \epsilon)$ of radius $\epsilon > 0$ and using the Wronskian \neref{eq:Wronskian}, we determine the value of $C_k$ to be
\begin{equation}
C_k = -\frac{\Omega_{n-1}}{2k + n - 2}.
\label{eq:constantForFinalSolution}
\end{equation}
Combining this with \eref{eq:Green'sFunctionForm}, we obtain the Green's function
\begin{equation}
G(\vec{r}, \vec{r}_p) = -\Omega_{n-1}\sum_{k \geq 0}\sum_{l = 1}^{\Gamma_{k}} \frac{R_k^{(\alpha)}(r_>)R_k^{(\bar{\alpha})}(r_<) \hat{Y}_k^l(\vec{\varphi}_p)Y_k^l(\vec{\varphi})}{2k + n - 2}.
\label{eq:GreensFunctionFinalTwo}
\end{equation}

\section{Multipole Hair and Discussion}
\label{sec:Results}

Let $q$ be the globally conserved Noether charge that results from integrating the Noether current $\nabla_\nu j^\nu = 0$ over the ST manifold. Since $j^0$ is the only nontrivial component of the vector current, Stokes' theorem \cite{Thorne} implies
\begin{equation}
q = \int j^0(r,\vec{\varphi}) \sqrt{|\det g|}\,\d r\, \d\vec{\varphi}= \int j^0(r,\vec{\varphi}) r^{n-1}\,\d r\, \d \vec{\Omega}_{n-1}.
\label{eq:conservedCharge}
\end{equation}
In our model, we consider the effect of a point source of strength $q$ located at $\vec{r}_p = (r_p, \vec{\varphi}_p)$, where $r_p > r_s$ (outside the event horizon). The normalizations \nerefs{eq:normalizationOne}{eq:normalizationTwo} establish that the physical source is then $j^0(r,\vec{\varphi}) = q\delta^n(\vec{r} - \vec{r}_p)$. Consequently, the solution $\Psi(\vec{r},\vec{r}_p)$ to \eref{eq:Poisson} that behaves appropriately is $\Psi(\vec{r},\vec{r}_p) = q G(\vec{r}, \vec{r}_p)$.

In this analysis, we shall examine the behavior of the field $\Psi$ at points $\vec{r} = (r,\vec{\varphi})$ for which $r > r_p$ as $r_p \rightarrow r_s^+$. Physically, this limit corresponds to a ``slow fall'' of the charge into the event horizon of the ST black hole. We assume the fall is slow enough such that our static considerations remain valid. In the following, the multipole moments are identified relative to the monopole term, which in $n+1$ dimensional spacetime is asymptotic to $r^{-(n-2)}$. Accordingly, terms asymptotic to $r^{- (\mu_k + n - 2)}$ characterize the $\mu_k$-pole moment.

For the $n = 3$ case, Cohen and Wald \cite{Cohen} found that the spacetime approaches the Reissner-Nordstr\"om geometry for any observer outside the event horizon, and hence that all electric multipole moments vanish, with the exception of the conserved monopole charge $q$. The conclusion is markedly different when $n > 3$. Here, the final ST black hole exhibits countably infinite nonzero multipole moments. Furthermore, there exist spatial dimensions $n > 3$ in which a nonzero number of the multipole moments are of infinite strength. These conclusions follow immediately from the Green's function \neref{eq:GreensFunctionFinalTwo}, but we shall prove them explicitly below. In doing so, we frequently reference the set
\begin{equation}
\LLambda \defeq \left\{k\in \Z^+ : (n-2) \nmid k\right\} \cup \left\{0\right\}.
\label{eq:LambdaSet}
\end{equation}
As $n\geq3$, $\LLambda = \{0\}$ if and only if $n = 3$.

We now consider the effect of lowering an electrostatic point charge of strength $q$ into a ST black hole. \eref{eq:vanishingOfEMMultipoles} implies $R_k^{(\bar{\alpha})}(r_s)$ is nonzero if and only if $k \in \LLambda$. Therefore, an observer outside the horizon at $\vec{r} = (r,\vec{\varphi})$ measures the field
\begin{equation}
\Psi(\vec{r}, \vec{r}_p)\big|_{r_p = r_s} = -q\Omega_{n-1}\sum_{k \in \LLambda}\sum_{l=1}^{\Gamma_k}\sum_{m\geq0}\frac{\tilde{\alpha}_{k,m}R_k^{(\bar{\alpha})}(r_s) \hat{Y}_k^l(\vec{\varphi}_p)Y_k^l(\vec{\varphi})}{r_s^{k + n-2}(2k + n - 2)}\left(\frac{r_s}{r}\right)^{k + (m+1)(n-2)}.
\label{eq:EMFieldFinal}
\end{equation}
If $n = 3$, then $\LLambda = \{0\}$. Furthermore, $\tilde{\alpha}_{0,m} = 0$ for all $m \in \Z^+$. Hence, in this case, $\Psi$ only has a monopole term. This agrees with Cohen and Wald's result \cite{Cohen}: the multipole moments of the field for an electrostatic point charge (except the monopole) vanish as the charge approaches the event horizon of a Schwarzschild black hole. If $n > 3$, however, then there exists a $k \in \LLambda \backslash \{0\}$ for which $R_k^{(\bar{\alpha})}(r_s) \neq 0$. As $\tilde{\alpha}_{k,0} \neq 0$ for $k > 0$, $\Psi$ has a $k$-pole moment. But $\tilde{\alpha}_{k,m} \neq 0$ for $k > 0$ and all $m \in \Z^*$. Therefore, the existence of a single $k$-pole moment (excluding the monopole) implies the existence of countably infinite multipole moments --- namely, all $\mu_k$-pole moments for which $\mu_k$ and $k$ are congruent modulo $n - 2$. This suggests that ST black holes acquire countably infinite electric multipole moments from infalling, electrically-charged matter. 

Interestingly, if there exists a $k \in \Z^+$ with divisor $d_k$ such that $n = 2(d_k + 1)$, which is true if and only if $n > 3$ is even, then, by \eref{eq:vanishingOfEMMultipoles} and the analysis thereafter, there exists a $k$-pole moment (and hence a countably infinite set of $\mu_k$-pole moments) of infinite strength. Therefore, in even dimensions $n > 3$, $\Psi$ diverges globally (i.e., is everywhere infinite). However, if $n > 3$ is odd, then all nonzero multipole moments, and hence $\Psi$, are everywhere finite.

The behavior of the field at the horizon as the source approaches the horizon can be determined by swapping $\vec{r}$ and $\vec{r}_p$ in \eref{eq:EMFieldFinal} and taking the limit $r_p \rightarrow r_s^+$. It is easy to see using \erefs{eq:schwarzschildTwo}{eq:schwarzschildOne} that the field is infinite at the horizon if $n > 3$ is even. Otherwise, the field is well-behaved and finite at the horizon. This divergence in even dimensions brings about the possibility of destruction of the horizon.

The conclusion that the final state of the ST black hole possesses countably infinite electric multipole moments presents a paradox. We expect the final state to be RNT in nature due to the uniqueness of the RNT solution among all non-degenerate, topologically hyperspherical, static, asymptotically-flat, electrovac solutions to the Einstein-Maxwell equations. However, the RNT black hole is hyperspherically symmetric, so it cannot possess electric multipole anisotropies. We conclude that the final state is not RNT in nature. In particular, the final state is not a non-degenerate, topologically hyperspherical, static, asymptotically-flat, electrovac solution to the Einstein-Maxwell equations. One (or more) of these characterizations must not apply to the final state, so to render it different from the RNT spacetime \footnote{Note that the final state necessarily obeys the Einstein-Maxwell equations since all results in this paper have derived from these equations.}.

Staticity and, at least for odd dimensions $n > 3$, asymptotic flatness can be assured, however. Staticity follows from the observation that our analysis never concerned itself with the rate at which the charge is lowered into the black hole. Thus, the lowering rate (the only non-static phenomenon in this study) can be assumed arbitrarily close to zero. For asymptotic flatness, we note that the source of the global divergence of $\Psi$ as $r_p \rightarrow r_s^+$ for even $n > 3$ is the factor of $R_k^{(\bar{\alpha})}(r_p)$ in \eref{eq:EMFieldFinal}. That this factor is independent of $r$ and $\vec{\varphi}$ implies $\partial_i \Psi$ diverges globally as $r_p \rightarrow r_s^+$ for even $n > 3$. Since the energy content of the field is related to $\partial_i\Psi$, the global divergence of $\partial_i\Psi$ suggests $\Psi$ may influence the asymptotic geometry. In particular, asymptotic flatness of the final state is not guaranteed for even $n > 3$. Conversely, for odd $n > 3$, $\partial_i \Psi$ is everywhere finite in the horizon limit. Asymptotic flatness can then be assured by merely tuning the strength $q$ of the charge to a value small enough (though nonzero) such that the geometry is unaffected by it. For odd dimensions, therefore, our starting assumption that the electric field does not influence the local spacetime geometry holds well as $r_p \rightarrow r_s^+$ for $q$ sufficiently small. The influence of $\Psi$ on the asymptotic geometry must then be particularly negligible, thereby preserving asymptotic flatness. At least for odd dimensions, these considerations guide us to the question of how a static and asymptotically-flat black hole can exhibit electric multipole fields.

Assuming the horizon of the final state is both non-degenerate and homeomorphic to $\Sphere^{n-1}$, then uniqueness of the RNT black hole forces the final state to be RNT spacetime. However, as we have remarked, RNT spacetime is hyperspherically symmetric, and the final state is not. As the final state (in odd dimensions) is static and asymptotically-flat, we are lead to the conclusion that one (or both) of the remaining assumptions (non-degenerate horizon and hyperspherical topology) is incorrect. If the horizon is degenerate, then the final state would be a counterexample to the expected non-degeneracy of static black hole solutions to the higher dimensional Einstein-Maxwell equations \cite{Rogatko1, Rogatko2}. Moreover, the final state would have a degenerate and necessarily ahyperspherical horizon in order to generate the multipole anisotropies. On the other hand, if the final state is not topologically hyperspherical, suggesting that infalling electric charges induce discontinuous topological transformations to the horizon, then the uniqueness of the RNT solution is invalidated. This allows for a topologically- and geometrically-ahyperspherical solution to characterize the final state, which is necessary for it to possess the multipole fields. While both these mechanisms ostensibly resolve the paradox (and are not immediately mutually exclusive), uncovering their exact details warrants further investigation.

We conclude with a comment on even dimensions $n > 3$. As, in this case, $\partial_i\Psi$ diverges in the horizon limit, the global spacetime geometry may be altered in a significant way. Thus, asymptotic flatness of the final state is not guaranteed. It follows that the non-degeneracy and/or hyperspherical topology of the horizon need not be violated (though are not immediately precluded from being violated) to generate the multipole anisotropies. This is because relaxing the assumption of asymptotic flatness is enough to render the ST solution non-unique among all possible non-degenerate, topologically hyperspherical, and static solutions to the Einstein equations \cite{Gibbons1, Gibbons2}. It is conceivable, therefore, that in the even dimensional case, the absence of asymptotic flatness in the final state accounts for the multipolar structure of the electric potential. Of course, as in the odd dimensional case, the exact details of this possibility require a more in-depth analysis, on which we hope to report soon.

\begin{acknowledgments}
The author is greatly indebted to T. A. Moore and B. Shuve for reviewing the present article, and to J. Gallicchio for discussions.
\end{acknowledgments}

\nocite{*}
\bibliography{STReferences.bib} 

\end{document}